\documentclass[11pt,twoside]{article}
\usepackage{./asp2010}

\resetcounters

\begin{document}

\title{Wind Models for Very Massive Stars in the Local Universe}
\author{Jorick S. Vink$^1$, J.M. Bestenlehner$^1$, G. Gr\"afener$^1$, A. de Koter$^2$, and N. Langer$^3$
\affil{$^1$Armagh Observatory, College Hill, Armagh, BT61 9DG, Northern Ireland, UK}
\affil{$^2$Astronomical Institute Anton Pannekoek, University of Amsterdam, Kruislaan 403, 1098 SJ, Amsterdam, The Netherlands}
\affil{$^3$Argelander-Institut f\"ur Astronomie der Universit\"at Bonn, Auf dem H\"ugel 71, 53121 Bonn, Germany}}
\begin{abstract}
Some studies have claimed the existence of a stellar upper-mass limit of 150$M_{\odot}$. 
A factor that is often overlooked concerns the issue that there might be a significant 
difference between the present-day and the initial mass of the most massive stars -- as 
a result of mass loss.
The upper-mass limit may be substantially higher, possibly exceeding 200$M_{\odot}$.
The issue of the upper mass-limit will however remain uncertain as long as there is only
limited quantitative knowledge of mass loss in close proximity to the
Eddington (=$\Gamma$) limit.
For this reason, we present mass-loss predictions from Monte Carlo radiative
transfer models for very massive stars up to 300$M_{\odot}$.
Using our new dynamical approach, we find an upturn or ``kink'' in
the mass-loss versus $\Gamma$ dependence, at the point where our model
winds become optically thick.
These are the first mass-loss predictions where the transition from 
optically {\it thin} O-star winds to optically {\it thick} Wolf-Rayet winds
has been resolved.
\end{abstract}

\section{Introduction}

Stellar winds from massive O and Wolf-Rayet (WR) stars are ubiquitous and 
form an important aspect of both stellar and galaxy evolution. 
Despite recent reductions in O-star mass-loss rates by factors of $\sim$3 in comparison 
to unclumped empirical rates, population synthesis models by Voss et al. (2009) 
-- utilizing rotating stellar models such as those of Meynet \& Maeder (2003) 
that employ theoretical Vink et al. (2000) rates that match the factors $\sim$3 reduced 
rates -- show that the role of stellar winds in massive-star feedback has {\it increased} over 
the last decade. The reason for this possibly counter-intuitive finding is that 
stellar rotation has increased luminosities, 
whilst the objects also enter the WR-phase sooner. 
Strong stellar winds may prevent the most massive stars from entering 
the eruptive Luminous Blue Variable (LBV) phase, at least at Galactic metallicity. 
It is these very massive stars (VMS) with their strong stellar winds 
that are thought to dominate the radiative and kinetic 
energy input in young ($<$10 Myr) clusters, as well as the radio-active 
26-Al input that gives rise to gamma-ray lines from Galactic star-forming 
regions like Orion and Carina (Voss et al. 2010, 2012). 

Our understanding of stellar mass loss from the most massive stars is 
also crucial for the determination of their final fates, with respect
to the formation of intermediate mass-black holes 
(Belkus et al. 2007; Yungelsen et al. 2008), the occurrence of pair-instability 
supernovae (Langer et al. 2007), and the mass distribution of neutron stars and 
black holes (Heger et al. 2003; Eldridge \& Vink 2006)

A most relevant issue concerns the stellar upper mass limit. 
Until recently, many studies accepted the existence of an upper limit 
of 120-150$M_{\odot}$, but a recent study by Crowther et al. (2010) claimed 
much higher luminosities and masses for the central WN5h objects of the R136 cluster 
than previous estimates (de Koter et al. 1997). 
One of the potential problems with the Crowther et al. luminosities is that 
the stars are clustered, and whilst Crowther et al. took great care in the assessment of a 
potential binary nature of the objects, there is a non-negligible chance that 
the stars are ``polluted'' with the light from line-of-sight objects.

In the context of the VLT Flames Tarantula Survey, we have recently 
discovered and analysed a new WN5h object, VFTS 682, a near-identical twin
of R136a3 (Evans et al. 2011, Bestenlehner et al. 2011). This object is in apparent 
isolation from the R136 cluster, some 30\,pc away, which raises 
interesting questions about its origin (see Bestenlehner et al. for a discussion on the 
object having formed in isolation or being a slow runaway). 
It also allowed us to assess the reliability of the Crowther et al. luminosity 
claims for the WN5h stars in the central cluster. 
Our finding of $\log L/L_{\odot} = 6.5 \pm 0.2$ 
for the ``isolated'' object VFTS\,682 seems to support the claim for high luminosities of 
WN5h stars in the Tarantula region.
As VFTS\,682 is in relative isolation from the 
dense cluster, we argue that the chance of a line-of-sight coincidence is much 
lower than for the WN5h stars in the core of R136, and we thus have independent evidence 
that the ``traditional'' 120 or 150$M_{\odot}$ stellar upper-mass limit has been broken.
The upshot is that mass-loss rates for stars with 
masses above 150$M_{\odot}$ are needed to establish the evolution and fate of the most 
massive stars. Here, we present predictions from 60 up to 300$M_{\odot}$.  
In the VMS regime, stars are identified as late-type nitrogen-rich WNh stars. 
These objects are extremely close to the Eddington limit $\Gamma~=~g_{\rm rad}/g_{\rm grav}~=~\kappa L/(4 \pi c G M)$.

\section{Mass loss predictions using the Monte Carlo method}

Stellar winds from massive stars are driven by radiation pressure  
as extensively described by e.g. Puls et al. (2008). 
Our Monte Carlo methodology is basically the same as the one used to predict 
mass-loss rates for standard OB stars (Vink et al. 2000).
It is based on the energy extraction method of Abbott \& Lucy (1985), and improved 
by de Koter et al. (1997) and Vink et al. (1999). 
The method underlying the mass-loss recipe however, was 
semi-empirical, in the sense that a $\beta$-type velocity law was assumed, that reached a certain 
terminal wind velocity $v_{\infty}$, which was consistent with empirical values (determined 
from blue edges of ultraviolet P\,Cygni profiles). 
As empirical $v_{\infty}$ values are quite accurate (to within $\sim$20\%), this was the preferred approach 
in order to obtain reliable theoretical mass-loss rates, that could serve as input 
in stellar evolution calculations. 
Nevertheless, now that we wish 
to predict mass-loss rates for VMS, we advance our approach, aiming 
to derive mass-loss rates without the use of empirical constraints.

M\"uller \& Vink (2008) suggested a new line force parametrization that 
explicitly depends on radius (rather than the velocity gradient as often assumed). 
Furthermore, a new iteration method was proposed, and utilized with respect to 
a 40 $M_{\odot}$ ``standard'' O star. The results 
were encouraging in that the predicted $v_{\infty}$ was only slightly ($\sim$10\%) larger than observed. 
More recently, Muijres et al. (2012) computed an extensive grid of 
mass-loss rates and terminal wind velocities, and tested the methodology of M\"uller \& Vink (2008) 
by comparison to hydrodynamical computations. Encouragingly, both methods give similar 
results for the supergiant regime, validating the 
M\"uller \& Vink (2008) approach for VMS.

Nugis \& Lamers (2002) highlighted the relevance 
of the iron-peak opacities in the deepest photospheric layers 
for the initiation of Wolf-Rayet winds, which was confirmed  
by the Gr\"afener \& Hamann (2008) 
models for hydrogen-rich late WNh stars.  
According to these wind models, the presence of opacity
peaks may cause $\Gamma$ to exceed unity in deep layers, leading to the formation 
of optically-thick winds.
In the Monte Carlo approach, we trace 
the radiative driving of the entire wind, and as 
most of the energy is transferred in the supersonic regime, we are less 
susceptible to the details of the photospheric region. 
Our strategy has the advantage that it 
allows us to explore the transition from transparent to dense stellar winds, 
that occurs in models that do include the full $\Gamma$ in subsonic regions 
for $\Gamma$ close to one.
As our Monte Carlo models capture the full physics in the layers around and above the sonic point, we argue 
that they correctly predict the qualitative behaviour of the transition from 
transparent to opaque winds (Vink et al. 2011 for further details). 

\begin{figure}
\includegraphics[width=12cm]{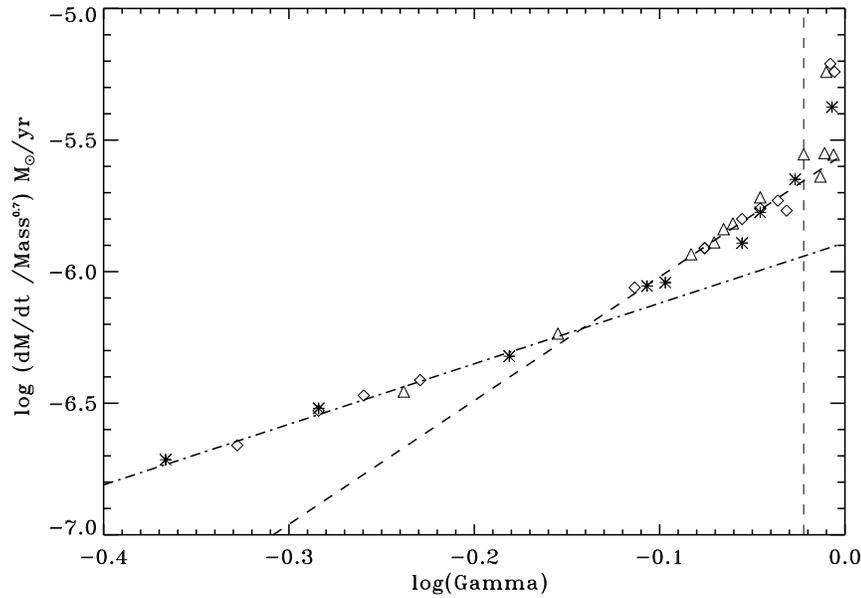}
\caption{Mass-loss predictions versus the Eddington parameter $\Gamma$ -- divided by $M^{0.7}$. 
Symbols correspond to models of different mass ranges (Vink et al. 2011).}
\label{f_mdotvink}
\end{figure}

\section{Resulting VMS mass-loss rates}

In Fig.~\ref{f_mdotvink} we present our new dynamically consistent 
mass-loss predictions for high $\Gamma$ values.
In order to assess the $\Gamma$ dependence separately from an additional
mass or luminosity dependence, we divided the new mass-loss rates 
by $M^{0.7}$.
For the lower $\Gamma$ range, the slope of $\dot{M}$ $\propto$ $\Gamma^{2.2}$ 
is in good agreement with previous O star results. We highlight the kink at  
$\Gamma$ $\sim$0.7. Above the kink, the winds are 
optically thick, and the $\dot{M}$ vs. $\Gamma$ relation has become much 
steeper, i.e. $\dot{M}$ $\propto$ $\Gamma^{4.7}$, in 
agreement with the earlier reported $\dot{M}$ $\propto$ $\Gamma^5$ for the semi-empirical (i.e. 
not necessarily locally dynamically consistent) calculations 
(Vink 2006). This steep slope in the high $\Gamma$ regime agrees well 
with the empirically determined slope in the Arches cluster stars, as well as that
determined using the PoWR models (see Gr\"afener et al. 2011 for further discussion).

In order to find out if radiation-driven mass-loss rates 
continue to rise with increasing $\Gamma$, we consider 
the mass-independent wind efficiency parameter $\eta = \dot{M} v_{\infty} / (L/c)$.  
At values of $\Gamma$ $\sim$0.5 we find $\eta$ numbers of $\sim$ 1, 
in accordance with standard Vink et al. (2000) models. 
However, when $\Gamma$ approaches unity, 
$\eta$ rises in a curved manner to values as high as $\eta$ $\simeq$3. 
Such large $\eta$ values are more commensurate with WR winds 
than those of common O stars, and these results thus confirm a natural extension from 
common O-type mass loss to more extreme WN behaviour.
Additionally, we note that the kink in the predicted mass-loss rate is accompanied by 
a transition of the velocity-law parameter $\beta$, with $\beta$ $\sim$1 around the kink, 
in accordance with models of M\"uller \& Vink (2008) 
and Muijres et al. (2012), to values larger than unity when $\Gamma$ 
exceeds 0.7. Large $\beta$ values had already been suggested to be  
more commensurate in WR stars previously, but it is reassuring 
to find that our modelling naturally predicts a transition in $\beta$ at the same 
point as where the kink in the mass-loss relations occurs. 

\begin{figure}
\includegraphics[width=12cm]{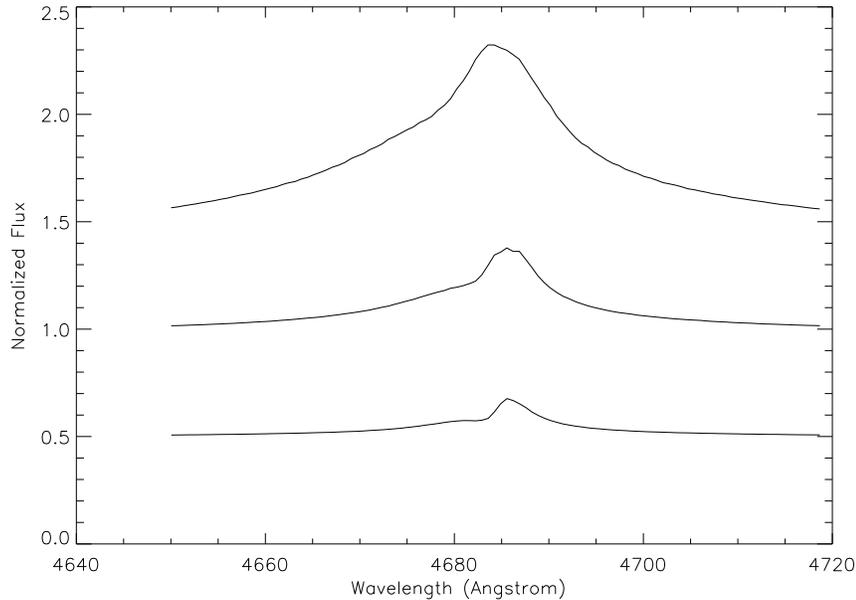}
\caption{The predicted normalized He\,{\sc ii}\,$\lambda 4686$ flux versus wavelength
for three values of $\Gamma$ (see Vink et al. 2011).}
\label{f_he}
\end{figure}

\section{Spectral morphology: the characteristic He 4686 Angstrom line}
\label{s_morph}

We have presented evidence for a 
transition in the mass-loss-$\Gamma$ exponent, as well as in the 
wind acceleration parameter $\beta$ 
from the moderate $\Gamma$ ``optically thin wind'' cases to 
``optically thick wind'' cases for objects that find themselves above 
$\Gamma$ $\ga$ 0.7, forming pseudo-photospheres. 
We might expect that the transition $\Gamma=$ 0.7 is the point where 
the spectral morphology of standard O-type stars changes from the common O and Of-types 
into a WN-type spectrum. The spectral sequence involving 
the Of/WN stars has a long history (e.g. Walborn et al. 1992)
but it had yet to be placed into 
a proper theoretical context. Figure~\ref{f_he} shows a sequence for the predicted 
He {\sc ii} 4686\AA\ lines for three gradually increasing values of $\Gamma$. 
Although the first spectrum below 
the transition $\Gamma$ already shows filled-in emission -- characteristic for Of stars -- the line-flux 
is rather modest compared to that found for the next two profiles 
with $\Gamma$ values exceeding the critical $\Gamma$ value.
The stars show very strong He {\sc ii} 4686\AA\ emission lines, more  
characteristic for WR stars of the nitrogen sequence (WN). 

\section{Summary and conclusions}

We presented mass-loss predictions from Monte Carlo radiative 
transfer models for VMS in the mass range 40-300$M_{\odot}$\ and 
with Eddington factors $\Gamma$\ in the range 0.4--1.0. 
Our models suggest that the spectral transition 
from Of to WN corresponds to a transition from relatively low $\Gamma$\ to 
high $\Gamma$ values (and larger velocity law parameter $\beta$) for WN stars.   
This assertion is not only based on the numerically larger mass-loss values, but 
more specifically on the finding of a ``kink'' at $\Gamma$ $=$ 0.7 
where the mass-loss slope in function of $\Gamma$ shows an upturn.

\end{document}